\documentclass[english]{emulateapj}
\usepackage{graphicx}
\usepackage{amssymb}

\makeatletter

\providecommand{\tabularnewline}{\\}

\makeatother

\begin{document}
\newcommand{\Mo}{M_{\odot}}
\newcommand{\Ro}{R_{\odot}}
\newcommand{\Lo}{L_{\odot}}
\newcommand{\SgrA}{\mathrm{Sgr\, A^{\star}}}
\newcommand{\Ms}{M_{\star}}
\newcommand{\Mbh}{M_{\bullet}}
\newcommand{\rMP}{r_{\mathrm{MP}}}
\newcommand{\aGW}{a_{\mathrm{GW}}}

\title{Massive perturbers and the efficient merger of binary massive black
holes}

\author{Hagai B. Perets and Tal Alexander\altaffilmark{1}}

\email{hagai.perets@weizmann.ac.il; tal.alexander@weizmann.ac.il}

\affil{Faculty of Physics, Weizmann Institute of Science, POB 26, Rehovot 76100, Israel} \altaffiltext{1}{The William Z. \& Eda Bess Novick career development chair}

\begin{abstract}
We show that dynamical relaxation in the aftermath of a galactic merger
and the ensuing formation and decay of a binary massive black hole
(MBH), are dominated by massive perturbers (MPs) such as giant molecular
clouds or clusters. MPs accelerate relaxation by orders of magnitude
relative to 2-body stellar relaxation alone, and efficiently scatter
stars into the binary MBH's orbit. The 3-body star--binary MBH interactions
shrink the binary MBH to the point where energy losses from the emission
of gravitational waves (GW) lead to rapid coalescence. We model this
process based on observed and simulated MP distributions and take
into account the decreased efficiency of the star-binary MBH interaction
due to acceleration in the galactic potential. We show that mergers
of gas-rich galactic nuclei lead to binary MBH coalescence well within
the Hubble time. Moreover, lower-mass binary MBHs ($<10^{8}\,\Mo$)
require only a few percent of the typical gas mass in a post-merger
nucleus to coalesce in a Hubble time. The fate of a binary MBH in
a gas poor galactic merger is less certain, although massive
stellar structures (e.g. clusters, stellar rings) could likewise lead
to efficient coalescence. These coalescence events are observable
by their strong GW emission. MPs thus increase the cosmic rate of
such GW events, lead to a higher mass deficit in the merged galactic
core and suppress the formation of triple MBH systems and the resulting
ejection of MBHs into intergalactic space. 
\end{abstract}

\keywords{black hole physics --- clusters --- galaxies: nuclei --- stars: kinematics
--- giant molecular clouds}

\section{Introduction}

There is compelling evidence that massive black holes (MBHs) exist
in the centers of most galaxies \citep{fer+00,geb+03,shi+03}. It
is believed that galaxies grow by successive mergers, during which
the two MBHs sink to the center of the newly formed galaxy by dynamical
friction and form a {}``hard'' binary MBH (BMBH) \citep{beg+80}
with a semi-major axis of \begin{equation}
a_{h}=[Q/(1+Q)^{2}]r_{h}(M_{12})/4\,,\label{e:a_h}\end{equation}
where $M_{12}\!=\! M_{1}\!+\! M_{2}$ is the mass of the binary, $Q\!\equiv\! M_{2}/M_{1}\!\le\!1$
is the mass ratio and $r_{h}(M_{12})$ is the radius of dynamical
influence of the BMBH %
\footnote{Defined here as the radius that encloses a stellar mass of $2M_{12}$
\citep{mer+06}. The threshold semi-major axis for a hard BMBH is
sometimes defined in terms of of $\sigma^{2}$, the typical velocity
dispersion in the center, $a_{h}\!=\! GM_{1}M_{2}\mu/4\sigma^{2}$,
where $\mu=M_{1}M_{2}/M_{12}$ is the reduced mass. However, this
is ill-defined since $\sigma^{2}$ usually varies with distance from
the BMBH.%
}, where typically, $a_{h}\!\sim\!1-10$ pc. After the BMBH hardens,
it continues to shrink by losing energy and angular momentum to stars,
which are ejected from the system (the {}``slingshot effect'', \citealt{sas+74})
or to gas with which it interacts dynamically. Once the separation
further decreases by 2--3 orders of magnitude, the BMBH rapidly decays
by the emission of gravitational waves (GWs) until the two MBHs coalesce. 

In mergers induced by interactions with stars, the merger timescale
depends on the rate at which new stars are supplied to BMBH-crossing
orbits ({}``loss-cone'' trajectories). Simulations show that this
supply rate is typically not high enough; the BMBH stalls before reaching
a small enough separation for efficient decay by GW emission, and
fails to coalesce in a Hubble time, $t_{H}$ (e.g. \citealt{ber+05},
see review by \citealt{mer+05b}). This result appears to contradict
the circumstantial evidence that most galactic nuclei contain only
a single MBH \citep{ber+06,mer+05b}, and furthermore, it implies
few such very strong GW sources, which future GW detectors, such as
the Laser Interferometric Space Antenna (LISA), hope to detect. 

Several mechanisms were suggested as means of accelerating BMBH coalescence
(see \citealt{mer+07} for a recent overview and discussion), either
involving interactions with stars ({}``dry mergers'') or with gas
({}``wet mergers''). These include re-ejection of stars that had
a previous interaction with the BMBH but were not ejected out of the
galactic core \citep{mil+03,ber+05}; BMBHs embedded in dense gas
\citep[e.g. ][]{iva+99,esc+05,dot+06b}; interactions of the BMBH
with a third MBH \citep{mak+94,bla+02,iwa+06}; accelerated BMBH coalescence
due to accelerated loss-cone replenishment in non-axisymmetric potential
\citep{yuq02,ber+06} or in a steep cusp \citep{zie06a,zie07}. It
is still unclear whether these mechanisms are efficient enough, or
whether they occur commonly enough to solve the stalling problem.
Efficient direct interaction of with gas requires the BMBH to be embedded
in a very dense massive central gas concentration. However, it is
unknown whether such amounts of gas exist there. For example, the
central $\sim\!2$ pc of the Galactic Center (GC) are gas-depleted
\citep{chr+05}, and other galaxies also show central gas cavities
in their nuclei \citep{sak+99}. The gas may also be dispersed by
the accreting BMBH before the merger is completed \citep{mer+05b},
and may not be efficient for minor mergers (\citealt{esc+05}; but
see \citealt{dot+06b}). It is likewise unknown whether the non-axisymmetric
potential assumed by the dry merger scenario of \citet{ber+06} is
generally present in the post-merger galaxy on the relevant scales.
Even if that is the case, actual demonstration of rapid BMBH coalescence
still awaits future $N$-body simulations with realistically high
$N$ \citep{ber+06}.

Here we explore another possibility, which is likely to apply in most
mergers: BMBH coalescence driven by massive perturbers (MPs), such
as giant molecular clouds (GMCs) or stellar clusters, in the post-merger
galaxy. MPs accelerate relaxation and scatter stars into the BMBH
orbit at high rates. Efficient relaxation by MPs was first suggested
by \citet{spi+51,spi+53} to explain stellar velocities in the galactic
disk. MPs remain an important component in modern models of galactic
disk heating (see e.g. \citealt{vil83,vil85,lac84,jen+90,han+02}
and references therein). A similar mechanism was proposed to explain
the spatial diffusion of stars in the inner Galactic bulge \citep{kim+01}.
In addition to dynamical heating, efficient relaxation by MPs was
suggested as a mechanism for loss cone replenishment and relaxation,
both in the context of scattering of Oort cloud comets to the Sun
\citep{hil81,bai83} and the scattering of stars to a MBH in a galactic
nucleus \citep{zha+02}. \citet{zha+02} also noted the possibility
of increased tidal disruption flares and accelerated MBH binary coalescence
due to MPs. Recently, \citet{per+07} (Paper I) studied in detail
MP-driven interactions of single and binary stars with a single MBH. 

In this study we apply the methods developed in Paper I to investigate
MP-driven interactions of stars with a BMBH, and the consequences
for BMBH coalescence. We explore different MP populations and merger
scenarios based on the available observations and simulations, and
estimate the BMBH coalescence rate. We also discuss additional implications:
the mass deficit in the galactic cores and the suppression of triple
MBH formation in galactic mergers. 

This paper is organized as follows. We begin with a general overview
of our calculations and the new results (\S \ref{s:overview}). In
\S \ref{s:MPlosscone} we briefly summarize the physics of MP-driven
loss-cone replenishment, which are derived in detail in Paper I. The
observations and theoretical predictions of MPs in the inner hundreds
pc of galactic nuclei are reviewed in \S \ref{s:MP_GC} and used
to construct the MP models used in our calculations (\S \ref{s:models})
. In \S \ref{s:merger_dyn} we briefly review the dynamics of BMBH
mergers; a detailed technical discussion is presented in appendices
\ref{a:stall} and \ref{a:energy}. We then present our procedure
for modeling the dynamical evolution of the BMBH coalescence under
various assumptions in \S\ref{s:merger_dyn} and analyze the results
of our calculations in \S \ref{s:Results}. We explore their implications
in \S \ref{s:Implications} and discuss and summarize our results
in \S \ref{s:summary}.

\section{Overview of the calculations and new results}

\label{s:overview}

This paper presents a first detailed study of the impact of MPs on
BMBH mergers. In this section we give an overview of the methods,
calculations and modeling that we use to study the dynamics of BMBH
mergers by MP-induced slingshot events, and to estimate the merger
timescale and the associated mass ejection from the nucleus. These
are described in detail in \S \ref{s:MPlosscone}--\S \ref{s:merger_dyn}
and in the appendices. We then briefly list the new results derived
here, and their implications. These are described and discussed in
detail in \S \ref{s:Results}--\S \ref{s:Implications}.

MPs shorten the BMBH merger timescale by accelerating the supply rate
of stars to the loss-cone (see \S \ref{s:MPlosscone}). The application
of loss-cone theory to the BMBH merger problem requires a model of
the MP properties: their mass, size, number and spatial distributions.
We review the available observations and numerical simulations of
MPs in merging galaxies of different morphological types (\S \ref{s:MP_GC},
tables \ref{t:MPs_prop} and \ref{t:MPs_abun}). We then construct
a few generic models for various types of merger scenarios, with and
without MPs, based on these data and on theoretical results (\S \ref{s:models},
table \ref{t:models}).

To evolve the BMBH in time in the context of a given merger scenario
(defined by a BMBH mass ratio and a MP model, see \S \ref{s:merger_dyn}),
we execute the following steps. (1) We establish the BMBH's initial
semi-major axis at the stalling point, where 2-body stellar relaxation
can no longer efficiently resupply the loss-cone. The stalling radius
is estimated semi-analytically by the formula derived in appendix
\ref{a:stall}, which are based on the $N$-body
simulations of \citet{mer06}. (2) At each time
step, we calculate the rate at which stars are scattered into the
loss-cone by MPs from all radii taking into account the BMBH's instantaneous
separation (Paper I; \S \ref{s:MPlosscone}).
(3) We calculate the hardness of the encounter (the ratio between
the encounter velocity and the BMBH orbital velocity) due to the acceleration
of the infalling star in the galactic potential, as derived in appendix
\ref{aa:galpot}. This effect, which was neglected in previous studies,
significantly decreases the energy extraction efficiency of the encounter.
(4) We estimate the average amount of BMBH orbital energy extracted
by the interaction with the stars scattered into the loss-cone. This
is based on numerical experiments of isolated 3-body encounters \citep[hereafter Q96]{qui96},
which are adapted in appendix \ref{aa:CH} to take into account the
changing hardness of the encounter in the course of the BMBH evolution.
(5) We include the energy extraction by GW emission, which increasingly
dominates the last stages of the merger. (6) We
update the BMBH semi-major axis according to the extracted energy
by both the slingshot effect and GW emission. (7) We end the dynamical
calculation when the emitted GW power exceeds that extracted by dynamical
interactions, and the BMBH enters the final rapid GW-induced decay
phase leading to coalescence. 

Our new results and their implications are as follows. (1) The inferred
properties and numbers of MPs in post-merger galactic nuclei are high
enough to effectively reduce the relaxation times there by orders
of magnitude. Consequently, BMBH decay due to MP-induced dynamical
interactions with stars is efficient enough in most mergers to lead
to coalescence within a Hubble time (typically within $10^{8-9}$
yrs), thereby solving the {}``last parsec'' problem. The rates of
GW emission from BMBH coalescence should thus trace the galactic merger
rates. (2) Since the MP-induced dynamical phase of BMBH mergers is
rapid, we predict that most BMBHs will be observed either electromagnetically
at separations larger than the hardening radius ($>\!1\,\mathrm{pc}$),
or by GW emission at much smaller separations (typically $\ll\!1$
pc), during the GW-dominated decay phase. (3) In most cases the BMBH
coalescence takes less than the typical time between major galactic
mergers. Therefore the formation of triple MBHs through multiple galactic
mergers is unlikely, and ejection of MBHs from unstable triples should
be rare. (4) The ejection of stars by the slingshot effect removes
mass from the galactic core ({}``mass deficit''). The more the BMBH
can decay, the larger the total ejected mass. The spatial scale of
the deficit reflects the origin of the scattered stars. We point out
that the magnitude and spatial extent of the deficit can be used to
discriminate between merger mechanisms. Specifically, the MP-induced
BMBH merger mechanism predicts a larger mass deficit over a larger
spatial extent than that due to stalled mergers by stellar relaxation
alone.

\section{Loss-cone refilling by massive perturbers}

\label{s:MPlosscone}

In Paper I we present a detailed quantitative analysis of the MP-induced
resupply of stars to nearly radial ,{}``loss-cone'' orbits, which
bring them within some threshold distance $q$ from the central mass
(MBH or BMBH), where they undergo a strong interaction with it ({}``event'')
and are thereby removed (scattered or destroyed). We show that the
resupply rate by MPs is orders of magnitude faster than that by stellar
2-body relaxation alone. This translates to an accelerated rate of
close interactions with the central object. Here we present a brief
qualitative summary of the results of paper I and their implications.

A star with orbital energy in the range $(E,E\!+\!\mathrm{d}E)$ on
a nearly-radial loss-cone orbit with angular momentum $J\!<\! J_{lc}$
will reach the MBH and be removed in about a single dynamical (or
orbital) time $P(E)$. When the resupply rate of such stars is slower
than the rate at which they are removed ($\sim\!1/P$), the phase-space
region $(\mathrm{dE},J\!<\! J_{lc})$ is nearly empty of stars. Conversely,
when the resupply rate is higher than $1/P$, that phase-space region
is nearly full, and its phase-space distribution is nearly isotropic.
At that point the rate at which stars with energy $E$ interact with
the central mass reaches its maximal value; further scattering can
not increase the rate. When the resupply of stars is driven by stellar
2-body relaxation, tightly bound regions of phase space (high $E$,
small typical $r$) are empty, since the angular size of the loss-cone
is large, the period is short and removal is fast. Conversely, loosely
bound regions (low $E$, large typical $r$) are full. Most of the
contribution to the total resupply rate comes from stars with energies
near the transition between the empty and full loss-cone regimes,
where $P\!\sim\! t_{r}$, the relaxation time \citep{fra+76,lig+77}.
This simple picture changes when MPs dominate relaxation, because
their distribution does not necessarily follows that of the stars,
and their properties (mass, size) may change with distance from the
center.

MPs of mass $M_{p}$ and space density $n_{p}$ dominate dynamical
relaxation over scattering by stars of mass $M_{\star}$ and space
density $n_{\star}$, when the ratio of the 2nd moments of the mass
distributions satisfies $\mu_{2}\!\equiv\!\left.n_{p}M_{p}^{2}\right/n_{\star}M_{\star}^{2}\!\gtrsim\! t_{r}^{\star}/t_{r}^{\mathrm{MP}}\gg\!1$.
The quantity $\mu_{2}$ thus expresses the MP enhancement factor of
the relaxation timescale, up to the ratio of the Coulomb logarithms
for star-star and star-MP scatterings, which takes into account the
extended size of the MPs . This can be shown by considering first
close encounters at the {}``capture radius'' $r_{c}\!\sim\! GM_{p}/v^{2}$,
where $v$ is the typical relative velocity. The {}``$nv\Sigma$''
rate of such encounters with a test star is then $t_{r}^{-1}\!\sim\! nvr_{c}^{2}\!\propto\! n_{p}M_{p}^{2}/v^{3}$.
Integration over all encounter distances further decreases the relaxation
time by a Coulomb logarithm factor that depends on the size of the
system and, if $R_{p}\!>\! r_{c}$, also on $R_{p}$, the size of
the MP%
\footnote{For example, on a scale of $r\!\sim\!50$ pc where the velocity dispersion
is $\sigma\!\sim\!{\cal O}(100\,\mathrm{km\, s^{-1}})$ and for MPs
of typical size $R_{P}\!\sim\!5$ pc, $\Lambda_{\star}\!\sim\! r\sigma^{2}/G\Ms$,
$\Lambda_{\mathrm{MP}}\!\sim\! r/R_{p}$ and $\log\Lambda_{\star}/\log\Lambda_{\mathrm{MP}}\!\sim\!8$.%
}. The impact of fast, MP-induced relaxation on the rate at which stars
enter the loss-cone can be incorporated into standard loss-cone theory
(e.g. \citealt{lig+77,you77}) by replacing the relaxation time due
to scattering by stars with that due to MPs, with the only modification
being the separate treatment of rare scattering events.

The effect of MPs on the BMBH loss-cone depends on the typical distance
from the center and on the BMBH's stage of dynamical evolution. Far
from the BMBH, MPs fill the loss-cone and drive the resupply rate
to its maximal value, while closer to the BMBH, even MPs can not fill
the loss-cone, but they still increase the supply rate by a large
factor $\sim\!\mu_{2}$. Closer still, the tidal field of the BMBH
gradually limits the size and mass of the MPs to the point where stellar
relaxation re-emerges as the dominant relaxation mechanism (see Fig.
3 in Paper I). The demarcation between the full and empty loss-cone
regions changes with time as the BMBH decays, the size of the loss-cone
decreases and the empty loss-cone region contracts. Irrespective of
these details, the overall effect of MPs is to dramatically increase
the loss-cone resupply rate over what would be expected from slow
stellar 2-body relaxation alone. 

We calculate the loss-cone resupply rate with some simplifying approximations.
We assume a spherically symmetric distribution, the \emph{ansatz}
$E\leftrightarrow GM(<r)/2r$ and the Keplerian expression for the
orbital period $P(E)$. We also make the conservative assumption that
one class of massive perturbers dominates relaxation, and neglect
the contributions from other mass components (e.g. stars). The full
expressions for the supply rates of stars deflected from a typical
radius $r$ in the empty loss and full loss cone regimes (\S 2 in
paper I, ) can be respectively approximated by \begin{equation}
\frac{\mathrm{d}\Gamma_{e}}{\mathrm{d}\log r}\!\sim\!\frac{N_{\star}(<\! r)}{\log(r/q)t_{r}}\,,\qquad\frac{\mathrm{d}\Gamma_{f}}{\mathrm{d}\log r}\!\sim\!\frac{q}{r}\frac{N_{\star}(<\! r)}{P(r)}\,,\label{e:Gammaef}\end{equation}
where $N_{\star}(<\! r)$ is the number of stars enclosed within $r$,
and where the dependence on the MPs properties enters through the
relaxation time, $t_{r}\!=\! t_{r}^{\mathrm{MP}}\!\propto\!\mu_{2}t_{r}^{\star}$.
The total loss-cone refilling rate is then given by integrating over
the contributions from all radii, taking into account the fact that
the refilling rate does not exceed the full-loss cone rate,\begin{equation}
\Gamma(q)=\int\frac{\mathrm{d}\Gamma(q)}{\mathrm{d}r}\mathrm{d}r=\int\min\left(\frac{\mathrm{d}\Gamma_{e}(q)}{\mathrm{d}r},\frac{\mathrm{d}\Gamma_{f}(q)}{\mathrm{d}r}\right)\mathrm{d}r\,.\label{e:Gammatot}\end{equation}
In the context of BMBH decay, the rate (Eq. \ref{e:Gammatot}) implicitly
depends on time through the evolution of $q\!=\! a(t)$, and the rate,
in turn, affects the evolution of $a$ through the BMBH evolution
equation (Eq. \ref{e:adot_dyn} below). The BMBH is evolved in time
by iterative numerical integration of the coupled rate and evolution
equations. 

The effects of MPs are significant in situations where perturbations
by stars alone are not efficient enough to refill the loss-cone on
the orbital timescale. MPs are most effective for large-$q$ processes,
such as close interactions between binaries and a single MBH (where
$q$ is the tidal disruption radius of the binary), or interactions
of single stars with a BMBH (where $q$ is the BMBH's semi-major axis).
This is because the larger $q$, the lower the empty loss-cone refilling
rate relative to that needed to completely fill the loss-cone (i.e.
$\Gamma_{e}/\Gamma_{f}\!\propto\!(r/q)/\log(r/q)$ is a decreasing
function of $q$ for $r/q\!\gg\!1$, Eq. \ref{e:Gammaef}).

\section{Massive perturbers in galactic nuclei}

\label{s:MP_GC}

The space density of MPs is much smaller than that of stars, so to
dominate relaxation ($\mu_{2}\!\gg\!1$) they must be significantly
more massive. Here we consider only MPs with masses $M_{p}\!\ge\!10^{4}M_{\odot}$,
such as stellar clusters and GMCs gas clumps. Intermediate mass black
holes (IMBHs) could be very effective MPs, but these are not considered
here since it is still unclear whether they actually exist. A summary
of the observed properties of MPs and those derived from simulations
is presented in tables \ref{t:MPs_prop} and \ref{t:MPs_abun}.

The lifespan of single MPs is affected by many factors, such as collisions
with other MPs; the central galactic tidal field, which can disrupt
them if they approach the center on an eccentric orbit or sink there
by dynamical friction; GMCs can also be dispersed by strong radiation
from stars and by supernovae. Thus, individual MPs are not expected
to survive much longer than a few local dynamical times, a time substantially
shorter than the galactic merger timescale (roughly the galactic dynamical
time), which in turn may be much shorter than the time required for
BMBH merger. However, MPs do not have to exist individually much longer
than a local dynamical time to have a strong effect on relaxation.
It is sufficient that the overall MP population maintains a steady
state, by the continuous formation or supply of new MPs to replace
those destroyed. There is both observational and theoretical support
for the long-term persistence of large populations of MPs in galactic
nuclei.

Observation of the GC indicate the steady state existence of a large
population of short-lived MPs over a Hubble time. As reviewed in detail
in Paper I, observations of dense gas in the central $\sim\!100$
pc of the GC reveal $N_{p}\!\sim\!100$ GMCs with a typical mass of
$\left\langle M_{p}\right\rangle \!\sim\!10^{5}\,\Mo$ (table \ref{t:MPs_prop})
\citep{oka+01}. GMCs and young stellar clusters are stages in the
path of star formation, and so the star formation rate can be used
to estimate the MP mass supply rate. \citet{fig+04} show that the
star formation history in the central projected 30 pc of the GC is
well described by continuous star formation over 10 Gyr at a rate
of $0.02\,\Mo\,\mathrm{yr^{-1}}$. Extrapolated out to 100 pc in the
$n_{\star}\!\sim\! r^{-2}$ stellar distribution of the inner bulge,
this corresponds to $\mathrm{d}\Ms/\mathrm{d}t\!\sim\!0.05\,\Mo\,\mathrm{yr^{-1}}$.
Since the mean star formation efficiency (fraction of mass turned
into stars) is on average very low, $f_{\star}\!\sim\!\mathrm{few}\times0.01$
\citep{mye+86}, the star formation rate translates to an MP mass
supply rate of $\mathrm{d}M/\mathrm{d}t\!\sim\!\mathrm{(d}\Ms/\mathrm{d}t)/f_{\star}\!\sim\!{\cal O}(1\,\Mo\,\mathrm{yr^{-1}})$,
and MPs formation or supply rate of \begin{equation}
\Gamma_{p}\!\sim\!\frac{\mathrm{d}M/\mathrm{d}t}{\left\langle M_{p}\right\rangle }\!\sim\!5\!\times\!10^{-5}\mathrm{yr^{-1}}\left(\frac{f_{\star}}{0.01}\right)^{-1}\!\left(\frac{\left\langle M_{p}\right\rangle }{10^{5}\, M_{\odot}}\right)^{-1}\,,\end{equation}
The mean lifetime of such GMCs is then \begin{equation}
t_{p}\!\sim\!\frac{N_{p}}{\Gamma_{p}}\!\sim\!2\!\times\!10^{6}\,\mathrm{yr}\left(\frac{f_{\star}}{100}\right)\left(\frac{\left\langle M_{p}\right\rangle }{10^{5}\,\Mo}\right)\left(\frac{N_{p}}{100}\right)\,,\end{equation}
 which is comparable to the dynamical time scale at $\sim\!100$ pc.
We therefore conclude that the observed MPs in the GC, together with
the inferred star formation rate and history, are fully consistent
with such a steady state MP population persisting over a Hubble time. 

Simulations of gas in galactic nuclei also show highly inhomogeneous
quasi-steady state conditions, with a very broad mass spectrum extending
over $\sim\!7$ orders of magnitude \citep{wad01,wad+01}. The nuclear
MP population in the Galaxy is probably representative of that in
nuclei of other spiral galaxies, where similar amounts of dense gas
are observed \citep{kod+05,sak+99,you+91}. Furthermore, estimates
of the gas supply timescales for star bursts in other galaxies also
suggest continuous gas supply over $10^{8}-10^{9}$ yr \citep{lei01}. 

The MP contents in the nuclei of post-merger galaxies are expected
to be yet larger than in the Galaxy, which does not appear to have
undergone a recent major merger. This is indeed observed in ULIRGS
\citep{san+96}, in merging galaxies \citep{eva+02,sak+06,cul+07},
and also seen in simulations \citep{bar+92}.

We now turn to a detailed discussion of the observed and inferred
properties of MPs in galactic nuclei of different types. These form
the basis of the MP models used in our numerical analysis of BMBH
mergers.

\subsection{Massive perturbers in spiral galaxies}

\label{sec:spiral_MPs}

The $\sim\!100$ GMCs observed in the GC, with masses in the range
$10^{4}$--$10^{7}\,\Mo$, contain $\mathrm{few}\times0.01$ of the
total dynamical mass on the $\mathrm{few}\times100$ pc scale and
a $\mathrm{few}\times0.1$ in the central $\sim\!100$ pc (see Paper
I for an extended discussion of the properties of MPs in our GC).
In contrast, the central \textasciitilde{}2 pc of the GC contain negligible
amounts of gas. Single molecular clouds cannot be resolved in the
nuclei of other spiral galaxies, but the total fraction of gas and
its distribution are usually quite similar to those observed in the
GC (e.g. \citealt{sak+99,saw+04}; see review by \citealt{hen+91}).
CO observations show that the gas contains very dense large clumps
that account for up to $\lesssim\!50\%$ of the total gas contents
in these regions \citep{dow+93,dow+98}. 

In addition to GMCs, many globular clusters \citep{fri95,ash+98}
and open clusters may inspiral into, or form in the galactic nucleus
in the course of their evolution (e.g. \citealt{gne+99}). For example,
the Galaxy contains hundreds of $\sim\!10^{3}\, M_{\odot}$ open clusters
and $\mathrm{few}\!\times\!10^{5}M_{\odot}$ globular clusters \citep{mey+91,fri95}.
Many more are observed in other galaxies \citep{ash+98}. If some
of these clusters contain IMBHs, they will contribute to the MPs population
even after the disruption of the host cluster is disrupted \citep{ebi+01,mill+02},
and will sink all the way to the center.

\subsection{Massive perturbers in elliptical galaxies}

\label{ss:ellip_MPs}

The gas fraction in elliptical galaxies is typically $10-100$ times
smaller than in spiral galaxies \citep{rup+97,kna+99}. However, in
some elliptical galaxies it is comparable or even larger than that
in spirals. Such gas-rich ellipticals are thought to have been formed
recently in a merger of two late type galaxies (e.g. \citealt{wik+97}).
In particular, ultra-luminous infrared galaxies (ULIRGs, see review
by \citealt{san+96}) have extreme amounts of gas, $10-100$ times
more than in the Galaxy, and can have as much or more mass in gas
compared to the mass in stars. Elliptical galaxies may well be evolved
merger products, where most of the dense gas in the core formed stars
(e.g. \citealt{ben+99}). In that case, it is plausible that the main
type of MPs would be the stellar clusters that were born of the GMCs,
rather than the GMCs themselves. Observations of stellar rings and
disks in the cores of elliptical galaxies indeed suggest that present-day
stellar structures reflect earlier gaseous structures \citep{dow+98}.
This is also consistent with the fact that ellipticals have larger
numbers of globular clusters than spirals, and that mergers are associated
with the formation of massive clusters \citep{ash+98,zha+99,kra+05,lar06}.

\subsection{Formation of massive perturbers in galactic mergers }

\label{ss:merger_MPs}

Simulations of mergers of gas rich spirals indicate that $\gtrsim\!50\%$
of the total gas mass in both galaxies is driven into the central
$\mathrm{few}\times100$ pc of the newly formed galaxy \citep{bar+91,bar+96},
where it probably forms massive clumps. In mergers of two gas-poor
ellipticals, stellar clusters may play a similar role. Many of the
newly formed stellar clusters will probably survive in the merged
nucleus \citep{por+02a}. In addition, many old globular clusters
will fall directly into the nucleus in the course of the merger \citep{gne+06},
or sink in by dynamical friction \citep{cap93}. While most will probably
be disrupted (O. Gnedin, priv. comm.), a significant fraction could
survive \citep[e.g. simulations by ][]{mio+06}. This central accumulation
of young and old stellar cluster could significantly shorten the relaxation
time. Further simulations are needed to address these questions quantitatively.

\begin{table*}
\caption{\label{t:MPs_prop}Observed and simulated properties of massive perturbers}

\noindent 
\begin{centering}
\begin{tabular}{lcccclp{1.5in}}
\hline 
{\footnotesize MP type} & 
{\footnotesize $M_{p}$ ($M_{\odot}$) } & 
{\footnotesize Mass Profile} & 
{\footnotesize $\left\langle M_{p}^{2}\right\rangle ^{1/2}\,(M_{\odot})$ } & 
{\footnotesize $R_{p}$ (pc)} & 
{\footnotesize References}
\tabularnewline
\hline
 &  &  &  &  & \tabularnewline
{\footnotesize GMCs in the GC} & {\footnotesize $10^{4}-10^{8}$} & {\footnotesize Power law ($\beta=1.2$)} & {\footnotesize $4\!\times\!10^{5}$} & {\footnotesize 5} & {\footnotesize \citet{oka+01,gus+04};} \tabularnewline 
&  &  &  &  & \citet{per+07}\tabularnewline
{\footnotesize Young clusters in the GC} & {\footnotesize $10^{3}-10^{5}$} & {\footnotesize Power law ($\beta=1.2$)} & {\footnotesize $3\!\times\!10^{4}$} & {\footnotesize 1} & {\footnotesize \citet{fig+99,fig+02,mai+04};} \tabularnewline
&  &  &  &  & \citet{bor+05,per+07}\tabularnewline
{\footnotesize Globular clusters in the Galaxy} & {\footnotesize $10^{2.5}-10^{6.5}$} & {\footnotesize Log normal } & {\footnotesize $1.9\times10^{5}$} & {\footnotesize 5} & {\footnotesize \citet{man+91}}\tabularnewline
{\footnotesize Young clusters in galaxies} & {\footnotesize $10^{4.5}-10^{6.5}$} & {\footnotesize Power law ($\beta=2$)} & {\footnotesize $4.3\times10^{5}$} & {\footnotesize 3} & {\footnotesize \citet{zha+99,kra+05};}\tabularnewline &  &  &  & & \citet{lar06}\tabularnewline
\hline
\end{tabular}
\par
\end{centering}
\end{table*}

\begin{center}
\begin{table*}
\caption{\label{t:MPs_abun}Mass fraction of observed and predicted\protect \\
massive perturbers in galactic nuclei }

\noindent \begin{centering}
\begin{tabular}{llll}
\hline 
{\footnotesize Galaxy} & {\footnotesize MP type} & {\footnotesize $M_{p}^{tot}/M_{\mathrm{dyn}}$ } & {\footnotesize References}\tabularnewline
\hline
 &  &  & \tabularnewline
{\footnotesize Milky Way} & {\footnotesize GMCs} & $\mathrm{few\times}0.1$ & {\footnotesize \citet{oka+01,gus+04}}\tabularnewline
\multicolumn{1}{c}{} & {\footnotesize Clusters} & $10^{-4}$ & {\footnotesize \citet{fig+99,fig+02,mai+04,bor+05}}\tabularnewline
{\footnotesize Spirals} & {\footnotesize GMCs} & $0.1$--$0.3$ & {\footnotesize \citet{kod+05,gus+04,sak+99,you+91}}\tabularnewline
{\footnotesize Elliptical mergers (Sim.) } & {\footnotesize Clusters} & {\footnotesize 0.2$^{a}$} & \citep{li+04,pri+06}\tabularnewline
{\footnotesize Ellipticals (Obs.)} & {\footnotesize GMCs} & $10^{-3}$--$10^{-2}$ & {\footnotesize \citet{rup+97,kna+99}}\tabularnewline
{\footnotesize ULIRGs } & {\footnotesize GMCs} & $0.3$--$0.6$ & {\footnotesize \citet{san+96}}\tabularnewline
{\footnotesize Merger (Obs.)} & {\footnotesize GMCs} & $0.3$--$0.6$ & {\footnotesize \citet{eva+02,sak+06,cul+07}}\tabularnewline
{\footnotesize Merger (Sim.)} & {\footnotesize{} }{\footnotesize GMCs} & $0.3$--$0.6$ & {\footnotesize \citet{bar+92}}\tabularnewline
\hline
\multicolumn{4}{l}{{\footnotesize $^{a}$ Assumed, based on simulations. See text.}}\tabularnewline
\hline
\end{tabular}
\par\end{centering}
\end{table*}

\par\end{center}

\section{Modeling massive perturber-driven BMBH coalescence}

\label{s:models}

Based on the observations and simulations described above, we formulate
three representative merger scenarios that include MPs, and compare
them to a merger scenario where only stellar 2-body relaxation plays
a role. The model parameters are listed in table \ref{t:models}.

The major merger scenario consists of a $Q\!=\!1$ merger of two gas-rich
galaxies. It is assumed that the merger triggers a large gas inflow
to the center, increasing the amount of gas there to $\sim\!1/2$
of the total dynamical mass ($\sim\!5$ times more than presently
in the center of the Milky Way; the mass of the cold gas in post-merger
galaxies can be even higher, but we take into account only the densest
regions that correspond to the more massive MPs). It is further assumed
that the MPs are similar to massive GMCs in our GC, that they have
a power-law mass function, $\mathrm{d}N_{p}/\mathrm{d}M_{p}\propto M_{p}^{-\beta}$
with $\beta=1.2$ (see MP model GMC1 in Paper I for details), and
that their spatial distribution is isotropic%
\footnote{While the geometry of central molecular zone of the Galaxy is flattened,
its height of $\mathrm{few}\times10$ pc implies that it is nearly
isotropic of the scale of interest of $\sim100$ pc.%
}.

The minor merger scenario consists of a $Q\!=\!0.05$ merger between
a large, massive gas-rich galaxy and a much smaller galaxy, which
only slightly perturbs the large galaxy and triggers only a moderate
gas inflow to the center. It is assumed that the nuclear gas mass
is $\sim\!1/3$ of the total dynamical mass ($\sim\!1.5$ times more
than presently in the center of the Milky Way). The MP properties
are the same as in the major merger scenario.

In the elliptical merger scenario we attempt to model a $Q\!=\!1$
merger of two equal mass gas-poor elliptical galaxies. We assume that
the MPs are mostly stellar systems such as clusters or spiral structures.
Lacking secure observations, we model the MPs after results from simulations
\citep{li+04,pri+06}. These simulations show that both the total
cluster birth-rate and the massive cluster birth-rate peak at the
center of the galaxy \citep{li+04}. We assume that the MP mass fraction
is $0.2$ of the total dynamical mass and that the cluster mass function
is a power-law with $\beta=2$ for $10^{5}\,\Mo\le M_{p}\le10^{7}\,\Mo$,
following the results of \citet{pri+06}. 

Finally, we consider, for comparison, a model that assumes that relaxation
in the post-merger galaxy is due to stellar 2-body interactions only.

In our calculations we assume that the stellar distribution over the
entire relevant distance range can be approximated by a singular isothermal
stellar distribution \begin{equation}
\rho(r)=\frac{\sigma_{\infty}^{2}}{2\pi Gr^{2}}\,,\label{e:iso_dist}\end{equation}
where the velocity dispersion $\sigma_{\infty}$, and hence the normalization,
is determined by the empirical $\Mbh/\sigma$ relation \citep[e.g.][]{wan+04}.
The MP distribution is assumed to follow the stars, down to a minimal
radius $r_{\mathrm{MP}}$, where the MPs are destroyed either by the
central tidal field, the radiation of the accreting BMBH, or the outflows
associated with the accretion or star formation triggered by the merger.
The exact value of $r_{\mathrm{MP}}$ is uncertain, since the processes
involved in the destruction of the MPs are complex, and the inner
cusp may be flattened due to previous mergers \citep{mil+02a}. Here
it is assumed that $r_{\mathrm{MP}}\!=\!2r_{h}$ %
\footnote{Note that the $\Mbh/\sigma$ and $\Mbh/M_{b}$ relations ($M_{b}$
is the mass of the bulge, with typical length scale $r_{b}$) then
imply that $r_{\mathrm{MP}}\!\propto\! r_{b}$.%
}. In the GC, $r_{\mathrm{MP}}\!\sim\!4$--$8\,\mathrm{pc}$. At $r\!>\! r_{\mathrm{MP}}$
our assumed minimal GMC mass of $5\times10^{4}\,\Mo$ is consistent
with observations \citep{bac+99,oka+01}. This minimal radius is probably
a conservative estimate, since transient dense clumps and dense cluster
cores can survive even at smaller distances, as seen in observations
in our GC, \citep{chr+05}, and found in simulations, \citep{wad+01,por+03}. 

\begin{table*}
\caption{\label{t:models}Massive perturber models in galactic mergers}

\begin{centering}
{\footnotesize }\begin{tabular}{lccccrcc}
\hline 
\multicolumn{1}{l}{{\footnotesize Merger model}} & {\footnotesize $Q$} & {\footnotesize $r/r_{h}\,^{a}$ } & {\footnotesize $M_{p}^{tot}/M_{\mathrm{dyn}}^{tot}$} & {\footnotesize{} $M_{p}(M_{\odot})$} & {\footnotesize $\beta\,^{b}$} & {\footnotesize $R_{p}$ (pc)} & {\footnotesize $\mu_{2}\,^{c}$}\tabularnewline
\hline
 &  &  &  &  &  &  & \tabularnewline
{\footnotesize Major} & {\footnotesize 1} & {\footnotesize $2$--$30$ } & {\footnotesize $1/2$} & {\footnotesize $5\!\times\!10^{4}$--$1\!\times\!10^{7}$} & {\footnotesize $1.2$} & {\footnotesize 5} & {\footnotesize $3\!\times\!10^{7}$}\tabularnewline
{\footnotesize Minor} & {\footnotesize 0.05} & {\footnotesize $2$--$30$ } & {\footnotesize $1/3$} & {\footnotesize $5\!\times\!10^{4}$--$1\!\times\!10^{7}$} & {\footnotesize $1.2$} & {\footnotesize 5} & {\footnotesize $5\!\times\!10^{6}$}\tabularnewline
{\footnotesize Elliptical } & {\footnotesize 1} & {\footnotesize $2$--$30$ } & {\footnotesize $1/5$} & {\footnotesize $1\!\times\!10^{5}$--$1\!\times\!10^{7}$} & {\footnotesize $2$} & {\footnotesize 3} & {\footnotesize $5\!\times\!10^{5}$}\tabularnewline
{\footnotesize Stars} & {\footnotesize ---} & {\footnotesize $1$--$30$ } & {\footnotesize $1$} & {\footnotesize $1$} & {\footnotesize ---} & {\footnotesize $0$} & {\footnotesize 1}\tabularnewline
\hline
\multicolumn{8}{l}{{\footnotesize \rule{0em}{1.5em}$^{a}$ Spatial extent of the MP distribution, assuming $N_{p}(r)\!\propto\! r^{-2}$.}}\tabularnewline
\multicolumn{8}{l}{{\footnotesize \rule{0em}{1.5em}$^{b}$ Assuming
$\mathrm{d}N_{p}/\mathrm{d}M_{p}\!\propto\! M_{p}^{-\beta}$}}\tabularnewline
\multicolumn{8}{l}{{\footnotesize \rule{0em}{1.5em}$^{c}$}{\footnotesize{}
$\mu_{2}\!\equiv\! N_{p}\left\langle M_{p}^{2}\right\rangle \left/N_{\star}\left\langle M_{\star}^{2}\right\rangle \right.$,
where }{\footnotesize $\left\langle M^{2}\right\rangle \!=\!\int M^{2}(\mathrm{d}N/\mathrm{d}M)\mathrm{d}M/N$. }}\tabularnewline
\hline
\end{tabular}
\par\end{centering}
\end{table*}

\section{BMBH merger dynamics}

\label{s:merger_dyn}

{A BMBH merger progresses through three stages \citep[See][]{mer06}.
(1) Gradual decay by dynamical friction to the point where the separation
between the two MBHs is $r_{12}\!\sim\! r_{h}(M_{1}$). (2) Formation
of a bound Keplerian pair, when $r_{12}\!<\! r_{h}(M_{1})$, through
rapid decay, initially by dynamical friction on $M_{2}$ and later
by the slingshot effect. This is followed by a slow-down of the decay
when $a\!\sim\! a_{h}$ and stalling, unless the the loss-cone is
replenished by a process more efficient than diffusion due to 2-body
relaxation. (3) Ultimately, the BMBH orbital decay rate is dominated
by GW emission, leading to final coalescence. The operational definition
of the stalling separation $a_{s}$ at time $t_{s}$ is the point
where the decay rate sharply decreases. Typically $a_{s}\!\sim\!{\cal O}(a_{h})$
(see appendix} \ref{a:stall}{). }

The slingshot effect occurs when $q$, the periapse distance of the
star from the BMBH center of mass, is of the order of the BMBH semi-major
axis $a$. Such stars are ejected and lost from the system, either
directly or after several repeated interactions with the BMBH, and
on average extract energy $\Delta E(q)$ from the BMBH. The evolution
of the BMBH energy, or equivalently, the decrease in $a$, is given
by

\begin{equation}
\!\frac{\mathrm{d}}{\mathrm{d}t}\left(\frac{GM_{1}M_{2}}{2a}\right)\!=\!\int_{0}^{\infty}\!\frac{\mathrm{d}\Gamma}{\mathrm{d}q}\Delta E(q)\mathrm{d}q\equiv\Gamma(a)\left\langle \Delta E\right\rangle \!(a)\,,\label{e:adot_dyn}\end{equation}
where $\Gamma(a)$ is the supply rate of stars that approach the BMBH
on orbits with $q\!<\! a$, and $\left\langle \Delta E\right\rangle \!\propto\! a^{-1}$
is the appropriately weighted mean extracted energy (\citealt{mil+03,mer+05b};
see detailed discussion in appendix \ref{aa:CH}). It then follows
from Eq. (\ref{e:Gammaef}) that the dynamical decay rate in the two
regimes scales as $\dot{a}_{\mathrm{dyn}}\!\propto\!-a/\log(r/a)$
or $\propto\!-a^{2}$, respectively, so that in both cases the dynamical
hardening rate decreases as $a$ decreases. Note that in the hard
BMBH limit ($a\!\rightarrow\!0$), when the loss-cone is small and
therefore full, $\mathrm{d}(1/a)/\mathrm{d}t\!\sim\!\mathrm{const.}$
(Q96). 

When the BMBH separation becomes small enough, the orbital decay rate
due to GW emission, $\dot{a}_{\mathrm{GW}}$, becomes higher than
the dynamical decay rate. We conservatively assume circular BMBHs
(eccentric BMBHs coalesce faster in the GW emission dominated phase).
The decay rate on a circular orbit due to the emission of GW is \citep{pet64}
\begin{equation}
\dot{a}_{\mathrm{GW}}=-\frac{64}{5}\frac{G^{3}\mu M_{12}^{2}}{c^{5}a^{3}}\,,\label{e:adot_gw}\end{equation}
which increases as $a$ decreases. The time to decay to $a\!=\!0$
from an initial semi-major axis $a$ is \begin{equation}
t_{\mathrm{GW}}=\frac{5}{256}\frac{c^{5}}{G^{3}}\frac{a^{4}}{\mu M_{12}^{2}}\,,\label{e:t_gw}\end{equation}
Since $\dot{a}_{\mathrm{dyn}}$ decreases with $a$, while $\dot{a}_{\mathrm{GW}}$
increases, there exists a transition BMBH separation, $\aGW$, such
that $\dot{a}_{\mathrm{dyn}}(\aGW)\!=\!\dot{a}_{\mathrm{GW}}(\aGW)$.
Once the BMBH shrinks to $a_{\mathrm{GW}}$, the coalescence is inevitable
as long as $t_{\mathrm{GW}}(a_{\mathrm{GW}})\!<\! t_{H}$ and as long
as the BMBH remains unperturbed. The total time from the hardening
semi-major axis $a_{h}$ to the coalescence is then 

\begin{equation}
t_{c}=t_{\mathrm{dyn}}(a_{h}\rightarrow a_{\mathrm{GW}})+t_{\mathrm{GW}}(a_{\mathrm{GW}}\rightarrow0)\,.\label{e:equal_times}\end{equation}

The dynamical decay timescale $t_{\mathrm{dyn}}(a_{h}\rightarrow a_{\mathrm{GW}})$
is of the order of the time it takes the BMBH to intercept and interact
with stars whose total mass equals its own , $t_{\mathrm{dyn}}\sim M_{12}/[\Ms\Gamma(a_{\mathrm{GW}})]$,
where $\Gamma$ is evaluated at $a_{\mathrm{GW}}$, where the rate
is slowest%
\footnote{\label{ft:dEdt}Every star that passes near the binary MBH extracts
from it binding energy of order $\Ms\varepsilon_{12}$, where $\varepsilon_{12}\!=\! G\mu/2a$
is the specific energy of the BMBH, so that $\mathrm{d}E\!=\!-GM_{1}M_{2}/2a^{2}\mathrm{d}a\!=\!(\Ms G\mu/2a)\Gamma(a)\mathrm{d}t$.
Integrating between $a_{h}\!\gg\! a_{\mathrm{GW}}$ with $\Gamma(a)\!\sim\!\left|(a/r)N_{\star}(<\, r)/P(r)\right|_{r_{\mathrm{MP}}}$
(when the loss-cone is filled by MPs that orbit as close as $r_{\mathrm{MP}}$
from the MBH) yields $t_{\mathrm{dyn}}\!\simeq\! M_{12}/\Ms\Gamma(a_{\mathrm{GW}})$.%
}. This estimate neglects the possibility that a fraction of the stars
are not ejected from the loss-cone, but return to interact again with
the BMBH. This can further accelerate the decay, but is not enough
in itself to prevent stalling \citep{mil+03}.

In order to calculate the MP-induced coalescence time, it is necessary
to compute both the rate at which stars are scattered by MPs into
the BMBH, and the energy they extract from the BMBH. Based on results
from 3-body scattering experiments \citep{hil83,qui96,ses+06a,ses+07},
we assume that a star whose periapse distance $q$ from the BMBH's
center of mass is smaller than the BMBH separation $a$, interacts
strongly with the MBH and is then ejected out of the system. We omit
the possibility of re-ejection, and we neglect soft scattering events
($q\!>\! a$) since these are inefficient in extracting energy from
the BMBH (see \citealt{ses+06a} and appendix \ref{aa:CH}). We thus
obtain a conservative upper limit on the coalescence time. 

Beginning with a hard BMBH of separation $a(t\!=\!0)\!=\! a_{s}$
(appendix \ref{a:stall}), we define the time-dependent loss-cone
periapse as $q\!=\! a(t)$ and calculate the loss cone rate $\Gamma(q)$,
using the methods described in Paper I and summarized here in \S
\ref{s:overview}. We follow the evolution of the BMBH separation
by numerically integrating the evolution equation with small enough
time steps such that $\mathrm{d}a\!\ll\! a$ (see \citealt{mil+03}
and \citealt{ses+06a,ses+07} for a similar approach) until the orbital
decay is dominated by GW emission (Eq. \ref{e:equal_times}), which
is effectively the coalescence time $t_{c}$. In simplified notation,
the evolution equation (Eq. \ref{e:dlogadt}) is

\begin{equation}
\frac{\mathrm{d}\log a}{\mathrm{d}t}=-2\frac{M_{\star}}{M_{12}}\int\bar{C}(a,r)\frac{\mathrm{d}\Gamma(a)}{\mathrm{d}r}\mathrm{d}r\,,\label{e:s_evol}\end{equation}
where $d\Gamma/dr$ is the differential loss cone replenishment rate
and $\bar{C}$ is the mean value of the dimensionless energy, $C\!\equiv\!\left.M_{12}\Delta E\right/2M_{\star}E_{12}$,
exchanged between the scattered star and the BMBH (see detailed derivation
and numeric estimation in appendix \ref{aa:CH}; $C\!=\!1$ corresponds
to the case where the specific energy carried by the star equals twice
that of the BMBH). The quantity $\bar{C}(a,r)$ depends on the hardness
parameter of the encounter $\zeta\!\equiv\!\sigma(r)/V_{12}(a)$,
defined as the ratio between the typical initial velocity of the scattered
star far from the BMBH, $\sigma(r)$ and the orbital velocity of the
BMBH, $V_{12}\!=\!\sqrt{GM_{12}/a}$. An additional $r$-dependence
is introduced by the acceleration of the star toward the BMBH by galactic
potential, which increases the relative velocity between the BMBH
and the star at the point of encounter over what it would have been
if the star fell toward an isolated BMBH (see appendix \ref{aa:galpot})
and decreases the efficiency of the slingshot effect (Figure \ref{f:Ceff}).
This non-negligible effect, taken into account here, was neglected
in previous estimations of the BMBH coalescence times \citep{ses+06a,qui96}.

\section{Results }

\label{s:Results}

\begin{figure}
\begin{tabular}{c}
\includegraphics[clip,width=1\columnwidth,keepaspectratio]{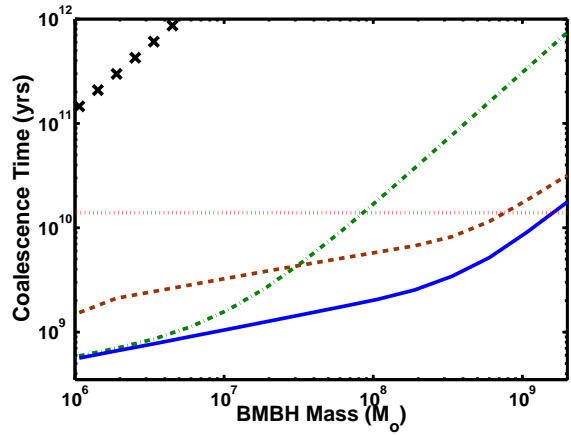}\tabularnewline
%\plotone{f1}\tabularnewline
\end{tabular}
\caption{\label{f:M-t_decay} The dynamical decay times, $t_{\mathrm{dyn}}$
of BMBHs from $a_{h}$ to $\aGW$ as function of the BMBH mass (the
time to final GW-induced decay, $t_{\mathrm{GW}}$, from $a_{\mathrm{GW}}$
to $a\!=\!0$ is negligible compared to the initial dynamical decay
phase). Different merger scenarios are shown (table \ref{t:models}):
major mergers (solid line), minor mergers (dashed line) and
elliptical mergers (dashed-dotted line). Without MPs (crosses, for
$Q\!=\!1$) the decay times are longer than the age of the universe
(horizontal line) for all BMBH mass in this range. }
\end{figure}

Figure (\ref{f:M-t_decay}) shows the total decay time of BMBHs in
the mass range $M_{12}\!=\!10^{6}$--$10^{9}$$M_{\odot}$ for different
merger scenarios. Stellar 2-body relaxation cannot replenish the loss
cone fast enough. In the absence of MPs, the merger proceeds in the
empty loss-cone regime, where the timescale is set by the slow relaxation
time (Eq. \ref{e:Gammaef}), leading to merger times orders of magnitude
longer than $t_{H}$. In contrast, when the MP number density is high
enough, or the loss-cone is small enough (lower BMBH mass), the loss-cone
is full, the merger time is the minimal possible, and is determined
by the size of the loss-cone and by the dynamical time (Eq. \ref{e:Gammaef}).
These conditions hold for gas-rich mergers across almost the entire
mass range ($M_{12}\!\lesssim\mathrm{few\times}10^{8}\,\Mo$), and
are also true for lower mass BMBHs in mergers of elliptical galaxies($M_{12}\!\lesssim\mathrm{few\times}10^{6}\,\Mo$).
For higher BMBH masses there are not enough MPs to refill the loss-cone.
However, the merger still evolves faster by a factor of $\sim\!\mu_{2}$
than it would with stellar relaxation alone (table \ref{t:models}),
until the BMBH separation decreases, the loss-cone is filled, and
the scattering rate reaches its maximal value. This fast MP-driven
evolution continues until the BMBH shrinks to the point where stellar
relaxation alone can fill the loss-cone. Since the BMBH spends most
of its time in those late stages, the overall decrease in the total
dynamical merger time, $t_{\mathrm{dyn}}$, is intermediate between
the maximal possible value of $\sim\!\mu_{2}^{-1}$ and that due to
stars alone (Eq. \ref{e:Gammatot}), $1\ll t_{\mathrm{dyn}}^{\mathrm{MP}}/t_{\mathrm{dyn}}^{\star}\ll\mu_{2}^{-1}$.

The results indicate that MPs drive rapid coalescence of BMBHs in
less than $t_{H}$ in most minor and major mergers. Moreover, for
most BMBHs coalescence occurs in less than a Gyr, which is comparable
to the dynamical timescale of the galactic merger itself \citep{bar+92}.
Our results indicate that massive BMBHs ($M_{12}\!\gtrsim\!10^{8}\,\Mo$)
in gas-poor ellipticals may take $\gg t_{H}$ to coalesce. However,
these estimates omit processes that could shorten the coalescence
time by an additional factor of a few, such as re-ejection of loss
cone stars \citep{mil+03,ber+05}. We find that MP-induced loss cone
refilling is effective in driving BMBHs of $M_{12}\!\sim\!10^{7}\,\Mo$
($10^{8}\,\Mo$, $10^{9}\,\Mo$) to coalescence in a Hubble time if
$0.005$ ($0.05$, $0.5$) of the total mass density in the galactic
nucleus is in clumped gas components with a mass function such as
observed in the GC (Figure \ref{f:M-t_decay}). Since the cores of
low-mass ellipticals with $M_{12}\!<\!10^{8}\,\Mo$ quite possibly
contain some clumped gas (up to $0.02$ of the MPs assumed in our
major merger model; table \ref{t:MPs_abun}). These are sufficient
for inducing rapid coalescence even in such systems, even if relaxation
stellar clusters or other coherent stellar structures is too slow. 

\begin{figure}
\begin{tabular}{c}
\includegraphics[clip,width=1\columnwidth,keepaspectratio]{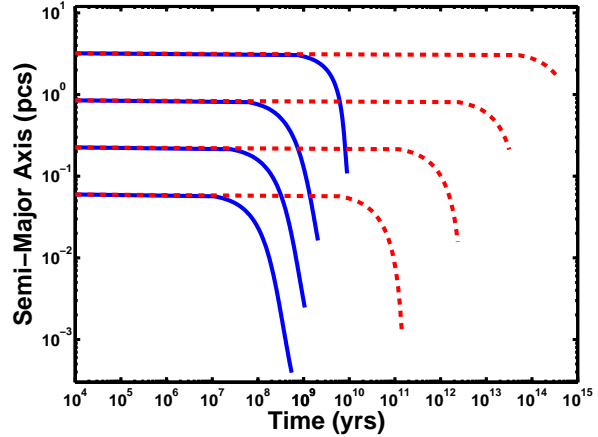}\tabularnewline
%\plotone{f2}\tabularnewline
\end{tabular}

\caption{\label{f:evolution} Evolution of the BMBH separation from $a_{s}$
to $a_{\mathrm{GW}}$ in a major merger due to 3-body scatterings
of stars. The evolution in the major merger scenario with MP-induced
relaxation (solid line) is compared to that with stellar relaxation
(dashed line) for BMBH masses of $10^{6},\,10^{7},\,10^{8}$ and $10^{9}$
$\Mo$ (from bottom up). }

\end{figure}

\begin{figure}
\includegraphics[width=1\columnwidth]{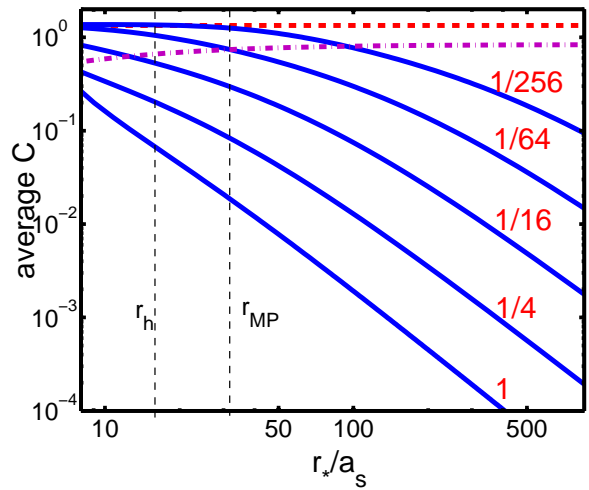}
%\plotone{f3}
\caption{\label{f:Ceff}The dependence of the mean dimensionless extracted
energy $\bar{C}_{\mathrm{eff}}\!=\!\bar{C}(Q,\zeta_{\mathrm{eff}})$
(Eq. \ref{e:eCmaxwell}) for $Q\!=\!1$, on the point of origin of
the deflected star, $r_{\star}/a_{s}$, for different stages of the
BMBH evolution $a/a_{s}$ (indicated by the numbers adjacent to the
lines), taking into account the acceleration in the Galactic potential.
The dashed horizontal line at the top is the asymptotic value of $\bar{C}_{\mathrm{eff}}$
in the hard limit ($a/a_{s}\!\rightarrow\!0$). The dash-dotted line
is the value of $\bar{C}$ for the case $a/a_{s}\!=\!1$, when the
Galactic potential is neglected. The vertical lines indicate the BMBH's
radius of dynamical influence $r_{h}$ and the inner cutoff of the
MP distribution $r_{\mathrm{MP}}$. Most of the stars are deflected
toward the BMBH from $r_{\star}\gtrsim r_{\mathrm{MP}}$. }

\end{figure}

Figure (\ref{f:evolution}) shows the evolution of the BMBH separation
for $M_{12}\!=\!10^{6}$, $10^{7}$, $10^{8}$ and $10^{9}\, M_{\odot}$
in major mergers ($Q\!=\!1$), with and without MPs. The BMBH separation
is evolved up to the point where the decay is dominated by GW and
coalescence follows soon after (the transition criterion $\dot{a}_{dyn}\!=\!\dot{a}_{\mathrm{GW}}$
and Eq. \ref{e:adot_gw} imply that the evolution curves steepen sharply
beyond the transition point). The evolution of BMBHs with MP relaxation
exhibits a short initial stalled phase, where the initially large
loss-cone is empty even in the presence of MPs, followed by a phase
of rapid decay. It should be noted that the decay phase does not display
the $a\propto t^{-1}$ evolution of a hard BMBH, expected when $\bar{C}\!\simeq\!\mathrm{const}$.
The acceleration of the infalling stars in the Galactic potential
softens the encounter with the BMBH and substantially reduces the
energy extraction efficiency. Figure (\ref{f:Ceff}) shows this efficiency
strongly depends on both the distance from which stars are deflected
to the BMBH and the BMBH separation. It should be emphasized that
acceleration by the galactic potential cannot not be neglected, since
it substantially reduces the efficiency of any BMBH slingshot mechanism,
in particular those where the potential gradient is steep (e.g. \citealp{zie06a,zie07})
or those where stars are deflected to the MBH from very large distances
(e.g. \citealp{ber+06}).

\section{Implications of MP-induced BMBH coalescence}

\label{s:Implications}

\subsection{Observations of BMBHs}

\label{s:Observations}

BMBHs can be observed as resolved objects in the initial dynamical
friction stage, when $a>a_{h}$, and possibly also in the second dynamical
decay stage when $a_{\mathrm{GW}}\!<\! a\!<\! a_{h}$ (in particular
massive BMBHs, whose $a_{h}$ is large, Eq. \ref{e:a_h}). They may
be detected indirectly (see review by \citealt{kom06}), or by GW
emission in the last GW-driven decay stage, when $a\!\lesssim\! a_{\mathrm{GW}}$.
Efficient MP-driven BMBH mergers progress rapidly through the second
dynamical decay stage. Thus, a prediction of the MP merger scenario
is that observed BMBHs should fall into a bimodal distribution: those
with $a>a_{h}$ and those with $a\lesssim a_{\mathrm{GW}}$, where
$a_{\mathrm{GW}}\!\ll\! a_{h}$. In contrast, dynamical scenarios
that lead to stalling, such as relaxation by stars alone, imply that
BMBHs with $a\lesssim\! a_{h}$ should be common.

The few direct and indirect observations of BMBHs available today
are consistent with the predictions of the MP scenario. There are
two known resolved double active galactic MBHs, NGC6240 with $a\!=\!1.4$
kpc \citep{kom+03}, and 0402+379 with $a\!=\!7$ pc \citep{rod+06},
just outside its hardening separation $a_{h}\!\sim\!3.5$ pc for $M_{12}\!\sim\!1.5\times10^{8}\, M_{\odot}$.
X-ray shaped radio galaxies, double-double radio galaxies, with pairs
of co-aligned symmetric double-lobed radio-structures, and AGN with
semi-periodic light curves or double peaked emission lines, were suggested
as signatures of close ($a\!\ll\!1$ pc), or recently merged BMBHs
\citep{kom06}. 

The detection of the GW signal from coalescing BMBHs would constitute
direct evidence of such events. Our calculations show that for most
galaxy mergers, the BMBH would coalesce within $t_{H}$, and so the
BMBH coalescence rate should follow the galaxy merger rate. In that
case the cosmic rate of these GW events could be as high as $10^{2}\,\mathrm{yr^{-1}}$
\citep{hae94,ses+04,eno+04}.

\subsection{Triple MBHs and MBH ejection}

The galaxy merger rate in dense clusters may be high enough ($>10^{-9}\,\mathrm{yr}^{-1}$;
\citealt{mam06}) for a second merger to occur before the first BMBH
coalesces. This would result in the formation of an unstable triple
MBH system, which will eject one of the MBHs at high velocity \citep{sas+74}.
This scenario was suggested as a possible solution for the stalling
problem, as the third component may drive the BMBH to high eccentricities
and to much more rapid coalescence \citep{bla+02,iwa+06,hof+07}. 

Because MPs accelerate BMBH coalescence in most mergers, our results
indicate that triple MBH systems and high-velocity MBHs ejected by
the slingshot mechanism should be rare%
\footnote{Recent observations of an apparently host-less quasar \citep{mag+05}
were interpreted as an ejected MBH \citep{hae+06,hof+06a,hof+07},
but see \citet{mer+06b} for an opposing view.%
} (Fig. \ref{f:M-t_decay}). Because of the rapid BMBH decay, in those
cases where a triple MBH is formed, it is expected that it will be
hierarchical. This would typically lead to fast coalescence of the
inner BMBH \citep{mak+94,iwa+06}, followed by the MP-driven decay
and coalescence of the newly formed central MBH with the outer MBH. 

\begin{figure}
\noindent \begin{centering}
\includegraphics[width=1\columnwidth]{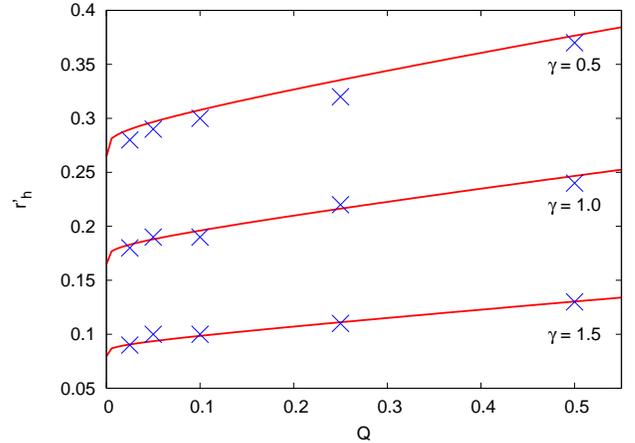}
\par\end{centering}

\caption{\label{f:rh_stall}The influence radius in the post-merger galaxy
at the stalling time, $r_{h}^{\prime}$, as function of $Q$, for
initial Dehnen density profiles with $\gamma=0.5,1.0,1.5$ (top to
bottom), as derived by \citet{mer06} in $N$-body simulations (crosses),
and by the approximate analytical expression Eq. (\ref{e:rhstall_approx})
(lines).}

\end{figure}

\subsection{Mass deficits}

The large number of stars ejected from the system during the BMBH
coalescence could change the stellar distribution of the BMBH environment.
It has been suggested that the mass deficit observed in some bright
elliptical galaxies is the result of such events \citep{mil+02,rav+02,gra04,fer+06}.
The total mass of ejected stars in the dynamical decay phase depends
only on the initial and final BMBH separations, 

\begin{equation}
M_{\mathrm{ej}}(t)\equiv\Ms\int_{0}^{t}\mathrm{d}t^{\prime}\int\mathrm{d}E{\cal F}(E,t^{\prime})\sim{\cal \mathcal{J}}M_{12}\ln\frac{a(0)}{a(t)}\,,\label{eq:m_lost}\end{equation}
where ${\cal F}$ is flux of stars supplied to the loss cone, and
$\mathcal{J}$ is a numerical factor approximately equal to $1/2\bar{C}$
(Q96; \citealt{mil+03,ses+07}). 

Previous studies of the mass deficit \citep{mil+02a,mer06} took into
account only the stars evacuated from the core before the BMBH stalled
at $a\!\sim\! a_{h}$ because of inefficient stellar relaxation. We
note that that there are between 2--7 further $e$-foldings between
$a_{h}$ and $a_{\mathrm{GW}}$ (Fig. \ref{f:evolution}). As a result,
when the BMBH merger is driven all the way to $a_{\mathrm{GW}}$,
the mass deficit will grow substantially (this was recently also noted
by \citealt{mer+07}). When the merger is driven by MPs, the mass
deficit will be on the $\sim(1$--$2)r_{\mathrm{MP}}$ scale, where
most of the scattered stars originate. We calculated the total mass
of stars that originated from such distances and that were ejected
from the core during coalescence. We found that these amount to approximately
30-40\% of the total stellar mass in these regions. 

We note that generally, the magnitude and spatial scale of the mass
deficit could, in principle, discriminate between different proposed
solutions for the stalling problem. In mergers driven by non-axisymmetric
potentials, most stars originate from large radii \citet{ber+06}
where the enclosed number of stars is very large. The fractional,
spatially-averaged mass deficit will therefore be very small, and
harder to detect. In contrast, scenarios that assume very steep cusps
\citep{zie06a,zie07} lead to substantial central mass depletion.
Even in gas induced mergers \citep{esc+05,dot+06b}, where stars play
a minor role, there may be an indirect mass deficit effect caused
by the inhibition of star formation due to heating of the gas by the
inspiraling BMBH.

\section{Discussion and Summary}

\label{s:summary}

We have shown that it is very likely that MPs play a dominant role
in the aftermath of galactic mergers. They shorten the relaxation
timescale in the galactic nuclei by orders of magnitudes relative
to 2-body stellar relaxation alone, and drive the newly formed BMBH
to rapid coalescence by the slingshot effect. The MP mechanism requires
only the existence of large enough inhomogeneities in the galactic
mass distribution. Since these occur naturally over a wide range of
conditions, the MP mechanism is robust and probably accelerates most
BMBHs mergers. The one possible exception may be mergers of two gas-poor
elliptical galaxies, where GMCs are less common. However, simulations
indicate that stellar clusters can play the role of MPs and drive
an efficient merger even in most of these cases. 

We conclude that most BMBHs are expected to coalesce within $t_{H}$,
even in cases where previous theoretical modeling, which did not consider
accelerated relaxation by MPs, predicts that the merging BMBH stalls.
This conclusion is based on conservative assumptions. We considered
only circular BMBHs, whereas eccentric BMBHs coalesce even faster
in the GW emission dominated phase, and we neglected the possible
concurrent effects of any of the other orbital decay mechanisms proposed
in the literature. It thus appears likely the BMBH coalescence is
in fact achieved on timescales $\ll\! t_{H},$ which implies that
BMBH coalescence GW events occur at the cosmic rate of galactic mergers.
Specifically, we predict that low-mass BMBHs, which are prospective
LISA targets, should coalesce within a merger dynamical time, $10^{8}$--$10^{9}$
yr. 

Efficient MP-driven BMBH coalescence have additional implications,
which were discussed here briefly. Fast BMBH mergers decrease the
probability of nuclei containing triple MBHs, and hence of ejected
MBHs, since in most cases, the BMBH coalescence time is shorter than
the mean time between galactic collisions. During the final stage
of the merger, when the BMBH separation shrinks from the hardening
radius to the final GW radius, a large number of stars will be ejected
from the nuclei. We find that this additional ejection stage could
appreciably increase the mass deficit of the newly formed nucleus,
beyond what is predicted taking into account only the earlier stages
of the merger \citep{mer06}. 

In summary, we have shown that the plausible existence of MPs in galactic
nuclei shortens the relaxation time by orders of magnitude. In particular,
MPs accelerate the dynamical decay of BMBHs by efficiently supplying
stars for the slingshot mechanism. This prevents stalling (the {}``last
parsec problem'') and allows the final coalescence of the BMBH by
GW emission within a Hubble time.

\acknowledgements{TA is supported by ISF grant 928/06, Minerva grant 8563 and a New
Faculty grant by Sir H. Djangoly, CBE, of London, UK. HP would like
to thank the Israeli Commercial \& Industrial Club for their support
through the Ilan Ramon scholarship.}

\appendix

\section{A. The stalling radius}

\label{a:stall}

This appendix presents a simple analytic approximation for the stalling
separation, $a_{s}$, {as function of the pre-merger
galactic density profile and the BMBH mass ratio $Q$}, which is based
on the $N$-body simulations of {\citet{mer06}.
Typically,} $a_{s}\!\sim\! a_{h}$ (Eq. \ref{e:a_h}), up to a factor
of a few. Assuming the \emph{ansatz} $a_{s}\rightarrow a_{h}$ in
the evaluation of the BMBH coalescence time can lead to inaccuracies
of up to a factor of a few, in particular for $Q\rightarrow1$.

{\citet{mer06} modeled typical galactic cores in
large $N$-body simulations of BMBH coalescence by Dehnen configurations
\citep{deh+93}, \begin{equation}
\rho=\frac{M}{[4\pi/(3-\gamma)]d^{3}}\frac{1}{(r/d)^{\gamma}(1+r/d)^{4-\gamma}}\,,\end{equation}
where $M$ is the total stellar mass, $d$ a scale length and $-\gamma$
the logarithmic slope at $r\!\ll\! d$. A central MBH of mass $M_{1}/M\!=0.01$
was added to the initial density distribution. We assume here that
the results derived for this particular class of models also apply,
at least approximately, to other initial density distributions and
MBH-to-stellar cluster mass ratios. }

{\citet{mer06} finds that the stalling radius can
be described to a good approximation, independently of $\gamma$,
by\begin{equation}
a_{s}=0.2Q/(1+Q)^{2}r_{h}^{\prime}(M_{12})=0.8\left[r_{h}^{\prime}(M_{12})/r_{h}(M_{12})\right]a_{h}\,,\label{e:a_stall}\end{equation}
where $r_{h}^{\prime}(M_{12})$ is the radius of influence of the
BMBH at the stalling time $t_{s}$, after} {the scouring
effect of the binary formation, which is estimated as follows. The
ejected mass at $t_{s}$ can be approximated analytically. \begin{equation}
\frac{\Delta M}{M_{12}}\simeq0.7Q^{0.2}\,.\end{equation}
The post-merger radius of influence $r_{h}^{\prime}$ can be estimated
to better than $3\%$ typically (Fig. \ref{f:rh_stall}), by assuming
that the post- density profile resembles the original profile, except
for the removal of $\Delta M$ from the center further out, so that
\begin{equation}
M(<r_{h}^{\prime})\!=\!2M_{12}^{\prime}\!\equiv\!{2M}_{12}\!+\!\Delta M\!=\! M_{1}(1+Q)(2+0.7Q^{0.2})\,.\end{equation}
The enclosed stellar mass in the initial Dehnen distribution is $M(<r)\!=\! M[r/(r+d)]^{3-\gamma}$,
and so $ $\begin{equation}
r_{h}^{\prime}(M_{12})/r_{h}(M_{12})=\left[\left(2M_{12}/M\right)^{1/(\gamma-3)}-1\right]\left/\left[\left(2M_{12}^{\prime}/M\right)^{1/(\gamma-3)}-1\right]\right.\,.\label{e:rhstall_approx}\end{equation}
The correction is thus a function of the inner cusp slope only. The
stalling separation (Eq. \ref{e:a_stall}) is a rising function of
$Q$. For $Q\!\rightarrow\!0$, $a_{s}\!\rightarrow\!0.8a_{h}$ irrespective
of $M_{12}/M$ or $\gamma$. For $Q\rightarrow1$, $M_{12}/M=0.01$
and $\gamma\!=\!2$, $a_{s}\!\rightarrow\!2.2a_{h}$.}

\section{B. BMBH energy extraction by interactions with stars}

\label{a:energy}

This appendix details how the mean BMBH energy that is extracted by
an encounter with a star from the galactic core is estimated using
results of isolated 3-body scattering experiments, which are available
in the literature.

\subsection{B.1 Adaptation of results from scattering experiments}

\label{aa:CH}

The rate at which the BMBH changes its binding energy $E_{12}=GM_{1}M_{2}/2a$
due to interaction of stars is 

\begin{equation}
\frac{\mathrm{d}E_{12}}{\mathrm{d}t}=\int_{0}^{\infty}\left\langle \Delta E(b;Q,\xi)\right\rangle 2\pi b\frac{\mathrm{d}\Gamma}{\mathrm{d}b}\mathrm{d}b\,,\label{e:dE}\end{equation}
where $b$ is the impact parameter, $\xi\!=\! v/V_{12}$ is the hardness
parameter ($v$ is the velocity of the incoming star at infinity relative
to the BMBH center of mass, ignoring the potential of the galaxy)
and $V_{12}\!=\!\sqrt{GM_{12}/a}$ is the BMBH's circular velocity
in the reduced mass system. The quantity $\left\langle \Delta E(b;Q,\xi)\right\rangle $
is the mean energy extracted by the star from the BMBH orbit averaged
over the BMBH orbital parameters, and $2\pi\mathrm{b(d}\Gamma/\mathrm{d}b)\mathrm{d}b$
is the rate at which stars are deflected into orbits with impact parameter
in the range $(b,b+\mathrm{d}b)$. In a \emph{Keplerian system} the
impact parameter $b\equiv xb_{0}$, with $b_{0}^{2}\!=\!2GM_{12}a/v^{2}\!=\!2a^{2}/\xi^{2}$
(Q96) is related to the periapse distance $r_{p}\!\equiv\! ya$ by
$b^{2}\!=\! r_{p}^{2}\left(1+2GM_{12}/r_{p}v^{2}\right)$, which can
be written as\begin{equation}
x^{2}=\xi^{2}y^{2}/2+y\,,\qquad y=\left.\left(\sqrt{1+2\xi^{2}x^{2}}-1\right)\right/\xi^{2}\,.\label{e:b}\end{equation}
 The extracted energy $\left\langle \Delta E\right\rangle $ is a
function of $r_{p}$, the mass ratio $Q$ and the hardness parameter
$\xi$. It was derived numerically by Monte Carlo simulations of 3-body
scattering (\citealt{hil83,qui96,ses+06a}), which show that it is
large and fairly constant for $r_{p}/a\lesssim0.5-2$, and then falls
rapidly to zero for $r_{p}/a\!\gtrsim\!0.5-2$ (see also \citealt{zie07}
for an extended discussion of the $r_{p}$-dependence of the extracted
energy). Unfortunately, the behavior of $\left\langle \Delta E(b;Q,\xi)\right\rangle $
as function of its parameters has only been partially documented.
\citet{hil83} studied the $b$ and $Q$ dependence only in the $\xi\rightarrow0$
limit, for specific values of the BMBH orbital eccentricities, \citet{ses+06a}
show plots only for $\xi\rightarrow0$, while Q96 explored the full
range of values for $\xi$ and $Q$ and averaged over the eccentricity,
but integrated over the $b$ dependence. For that reason, it is not
possible to use these results to evaluate Eq. (\ref{e:dE}) explicitly,
and it is necessary to resort to an approximate formulation. Following
the trends seen in the $\xi\rightarrow0$ simulation results, we adopt
here a step function approximation, which is based on the assumption
the $\left\langle \Delta E\right\rangle $ is roughly constant between
$b\!=\!0$ and an effective maximal impact parameter $b_{1}$, and
write \begin{equation}
\frac{\mathrm{d}E_{12}}{\mathrm{d}t}\sim\Delta\overline{E}_{1}\Gamma(b_{1})\,,\label{e:dEapprox}\end{equation}
where the $b$-averaged extracted energy is defined as \begin{equation}
\Delta\overline{E}_{1}\equiv\overline{\left\langle \Delta E(Q,\xi)\right\rangle }=\left.\int_{0}^{b_{1}}2\pi b\left\langle \Delta E(b;Q,\xi)\right\rangle \mathrm{d}b\right/\pi b_{1}^{2}\,,\label{e:dE1}\end{equation}
and where the total rate of stars with impact parameter $b\!<\! b_{1}$
is \begin{equation}
\Gamma(b_{1})=\int_{0}^{b_{1}}2\pi b\frac{\mathrm{d}\Gamma}{\mathrm{d}b}\mathrm{d}b\,.\end{equation}
It should be noted that the step function approximation implies that
the periapse-averaged energy extracted per star should not depend
strongly on the mode of loss-cone replenishment, whether it is in
the empty loss cone regime, where stars diffuse into the loss-cone
from its boundary ($r_{p}\!\sim\! a$), or whether it is in the full
loss cone regime, where the orbits span the entire range $0\!\le\! r_{p}\!\le\! a$.
This behavior is indeed seen indirectly in the simulations of \citet[Fig. 2]{mer+06},
where the mass ejection rate remains nearly constant as the system
transits from the full to the empty loss-cone regime. 

Q96 does not quote the simulation results in terms of $\left\langle \Delta E\right\rangle $
directly, but rather in terms of a related dimensionless quantity
$H_{1}$, which expresses the rate at which energy is extracted from
the BMBH by an ambient background of stars with mass density $\rho$
and velocity $v$ at infinity, $\mathrm{d}(1/a)/\mathrm{d}t=H_{1}G\rho/v$.
\citet{hil83} and Q96 express the extracted energy in a dimensionless
form,

\begin{equation}
C\!\equiv\!\left.M_{12}\Delta E\right/2M_{\star}E_{12}\!=\! a\Delta\varepsilon/G\mu\,,\end{equation}
where $\Delta\varepsilon\!=\!\Delta E/M_{\star}$ is the specific
energy extracted by the star and $\mu\!=\! M_{1}M_{2}/M_{12}$ is
the reduced mass ($C\!=\!1$ corresponds to the case where the specific
energy carried by the star equals twice that of the BMBH). In terms
of $C$, Eq. (\ref{e:dE1}) can be written as\begin{equation}
\overline{C}_{1}=\left.\int_{0}^{b_{1}}2\pi b\left\langle C(b;Q,\xi)\right\rangle \mathrm{d}b\right/\pi b_{1}^{2}\label{e:C1}\end{equation}
The quantity $H_{1}$ is related to the orbitally averaged $\left\langle C\right\rangle $
in terms of the dimensionless impact parameter $x$ by (Q96, Eq. 11)
\begin{equation}
H_{1}(Q,\xi)=8\pi\int_{0}^{\infty}x\left\langle C(x;Q,\xi)\right\rangle \mathrm{d}x=\left.4\pi\int2\pi b\left\langle C\right\rangle \mathrm{d}b\right/\pi b_{0}^{2}\,.\label{e:H1}\end{equation}
 A comparison of Eqs (\ref{e:C1}) and (\ref{e:H1}) shows that $\overline{C}_{1}$
is related to the values of $H_{1}$ quoted by Q96 through the relation
\begin{equation}
\overline{C}_{1}(Q,\xi)=H_{1}(Q,\xi)(b_{0}/b_{1})^{2}/4\pi\,.\end{equation}
Let $r_{p1}\!=\! y_{1}a$ be the periapse that corresponds to the
effective impact parameter $b_{1}$. It then follows that \begin{equation}
\overline{C}_{1}(Q,\xi)=\left.H_{1}(Q,\xi)\right/\left[4\pi y_{1}(1+y_{1}\xi^{2}/2)\right]\,.\label{e:avC1}\end{equation}

Q96 suggests an analytic fit to the behavior of $H_{1}$ as function
of $Q$ and $\xi$ (Q96, Eq. 16 and table 1), 

\begin{equation}
H_{1}(Q,\xi)=\frac{H_{0}(Q)}{\sqrt{1+\left[\xi/w_{0}(Q)\right]^{4}}}\,,\label{e:H1fit}\end{equation}
where here we use a slightly different notation from Q96, $w_{0}\!=\! w_{Q96}/V_{12}$,
so that the values of $w_{0}(Q)$ are the numeric values in the 3rd
column of table 1 in Q96, or given by the analytic fit formula (Q96,
Eq. 17) \begin{equation}
w_{0}(Q)\simeq0.85\sqrt{M_{2}/M_{12}}=0.85\sqrt{Q/(1+Q)}\,.\end{equation}
The tabulated values of $H_{0}(Q)$ (Q96, table 1, 2nd column) are
nearly independent of $Q$, with $\bar{H}_{0}\!=\!21.1_{-3.2}^{+1.4}$
over the range $Q\!=\!1/256$ to $Q\!=\!1$ (the lower values corresponding
to larger $Q$). 

For a  distribution of velocities characterized by a 1D velocity dispersion
$\sigma$, the effective $H$ is given in terms of $\zeta\!=\!\sigma/V_{12}$
by (Q96, Eq. 18) \begin{equation}
H(Q,\zeta)=H_{1}\left(Q,\sqrt{3}\zeta\right)\left\{ \sqrt{2/\pi}+\log\left[1+\alpha\left[\zeta/w_{0}(Q)\right]^{\beta}\right]\right\} \,,\end{equation}
 where $\alpha\!=\!1.16$ and $\beta\!=\!2.40$. This can then be
approximately related to the mean ejection energy as in Eq. (\ref{e:avC1})
by \begin{equation}
\overline{C}(Q,\zeta)=H(Q,\zeta)/\left[4\pi y_{1}(1+y_{1}\zeta^{2}/2)\right]\,.\label{e:eCmaxwell}\end{equation}
The nature of the approximation here is that the translation between
$b$ and $r_{p}$ is done for a representative velocity, assuming
Keplerian motion. For the purpose of numeric calculations we assume
$y_{1}\!=\!1$.

The relevant scale for MPs is $r_{\mathrm{MP}}\!\gtrsim\!2r_{h}$,
which encloses $>\!4M_{\bullet}$ in stars. On that scale the potential
is dominated by the stars. For a $r^{-2}$ stellar density distribution
far from the MBH, the velocity dispersion is $\sigma^{2}(r)\!\simeq\! GM_{\star}(<r)/r\!\simeq\!\mathrm{const}$.
Here we represent the typical initial stellar velocity by the circular
velocity $v\!=\!\sqrt{2}\sigma$. This velocity needs to be corrected
for the fact that the star is accelerated by the galactic potential
as it falls toward the MBH. In the fictitious 3-body system (the BMBH
and the star), its effective hardness parameter , $\zeta_{\mathrm{eff}}(r)\!>\!\zeta$,
depends on the star's point of origin (see \S \ref{aa:galpot}).
Thus, the BMBH total decay rate is given by integrating over the contribution
of stars originating from all radii, with the loss-cone size expressed
in terms of the periapse, \begin{equation}
\frac{\mathrm{d}\log a}{\mathrm{d}t}=-2\frac{M_{\star}}{M_{12}}\int\overline{C}\left[Q,\zeta_{\mathrm{eff}}(r;a)\right]\frac{\mathrm{d}\Gamma(<y_{1}a)}{\mathrm{d}r}\mathrm{d}r\,.\label{e:dlogadt}\end{equation}
 Note that the {}``hard limit'', $\mathrm{d}(1/a)/\mathrm{d}t\!=\!\mathrm{const}$,
is recovered when the cross-section for interacting with the BMBH
is dominated by the gravitational cross-section term ($\zeta\rightarrow0$,
$\overline{C}\!=\!\mathrm{const}$, Eq. \ref{e:eCmaxwell}), and $\Gamma\!\propto\! a$
(full loss-cone regime, Eq. \ref{e:Gammaef}).

\subsection{B.2 Dependence on the galactic potential}

\label{aa:galpot}

The extracted energy (Eq. \ref{e:eCmaxwell}) depends on the hardness
of the encounter $\zeta$. In the soft encounter limit ($\zeta\rightarrow0$),
the interaction with the BMBH is strong and independent of $\zeta$
(Eqs. \ref{e:avC1}, \ref{e:H1fit}). The hardness of the encounter
depends on the star's point of origin. The farther away it starts
from the center, the more it will be accelerated by the galactic potential,
and the faster it will be when it approaches the BMBH. Since MPs typically
deflect stars into the loss-cone from large distances, the effect
of the galactic potential in making the encounters harder and less
efficient cannot be neglected (Figure \ref{f:Ceff}).

The 3-body scattering experiments of \citet{qui96} took into account
only the potential of the BMBH. To relate the energy extraction to
that of a star falling in the combined potential of the BMBH and the
galaxy, it is necessary to calculate the corresponding effective initial
velocity the star should have in the fictitious 3-body system containing
only the BMBH and the star.

The gravitational potential in a spherical stellar system with mass
density $\rho(r)$ surrounding a central BMBH of mass $M_{12}$ is
\begin{equation}
\phi(r)=-4\pi G\left[\frac{1}{r}\int_{0}^{r}\rho(x)x^{2}\mathrm{d}x+\int_{r}^{\infty}\rho(x)x\mathrm{d}x\right]-\frac{GM_{12}}{r}\,.\end{equation}
For a power-law mass density profile between $r_{1}\ll r_{2}$, \begin{equation}
\rho(r)=\rho_{0}\left(\frac{r}{r_{0}}\right)^{-\alpha}\,,\end{equation}
the enclosed stellar mass is ($\alpha\neq3$, $r\!\gg\! r_{1}$) \begin{equation}
M(<r)\simeq\frac{4\pi}{3-\alpha}\rho_{0}r_{0}^{3}\left(\frac{r}{r_{0}}\right)^{3-\alpha}\,.\end{equation}
The potential at $r_{1}\!<\! r\!<r_{2}$ is\begin{equation}
\phi(r)=-\frac{GM_{12}m(r)}{r}-4\pi G\!\int_{r}^{\infty}\!\!\rho(x)x\mathrm{d}x=-\frac{GM_{12}m(r)}{r}-4\pi G\rho_{0}r_{0}^{\alpha}\left\{ \begin{array}{lr}
\ln(r_{2}/r)\,, & \alpha\!=\!2\\
\frac{1}{2-\alpha}\left(r_{2}^{2-\alpha}\!-\! r{}^{2-\alpha}\right)\,, & \alpha\!\neq\!2\end{array}\right.\,,\end{equation}
where $m(r)=1+M(<\! r)/M_{12}$ is the total mass up to radius $r$
relative to the BMBH mass.

Suppose a star starts falling toward the BMBH with velocity $v$ from
an initial radius $r$ down to a radius $r_{1}$ close enough to the
BMBH so that $M(<r_{1})\rightarrow0$. The specific orbital energy
of the star, $\varepsilon$, is conserved,\begin{equation}
\varepsilon=\frac{1}{2}v^{2}+\phi(r)=\frac{1}{2}v_{1}^{2}+\phi(r_{1})\,.\end{equation}

Taking the velocity of this star at $r_{1}$ as the velocity it would
have if it began falling from the same distance in the fictitious
3-body system, we can find what the \emph{effective} velocity, $v_{\mathrm{eff}}$,
it should have at $r$ far away from the BMBH \begin{equation}
\varepsilon'=\frac{1}{2}v_{\mathrm{eff}}^{2}+\phi'(r)=\frac{1}{2}v_{1}^{2}+\phi'(r_{1})\,,\end{equation}
where \begin{equation}
\phi'(r)=-\frac{GM_{12}}{r}\,.\end{equation}

The effective velocity it then \begin{equation}
v_{\mathrm{eff}}^{2}=v^{2}+2[\phi'(r_{1})-\phi'(r)+\phi(r)-\phi(r_{1})]=v^{2}+2GM_{12}\left[-\frac{1}{r_{1}}\!+\!\frac{1\!}{r}-\!\frac{m(r)}{r}\!+\!\frac{m(r_{1})}{r_{1}}\!+\!\frac{4\pi G}{M_{12}}\int_{r_{1}}^{r}\rho(x)x\mathrm{d}x\right]\,.\end{equation}
Since $m(r_{1})\rightarrow1$, \begin{equation}
v_{\mathrm{eff}}^{2}=v^{2}+2G\left[-\frac{M(<r)}{r}+4\pi\int_{r_{1}}^{r}\rho(x)\mathrm{d}x\right]\,.\end{equation}
For the $\alpha\!=\!2$ power law assumed here ($r\!\gg\! r_{1}$),\begin{equation}
v_{\mathrm{eff}}^{2}\simeq v^{2}+8\pi G\rho_{0}r_{0}^{2}\left[\ln\left(\frac{r}{r_{1}}\right)-1\right]\,.\end{equation}
The effective hardness parameter is then $\zeta_{\mathrm{eff}}(r)\!=\! v_{\mathrm{eff}}/V_{12}$,
which is used to evaluate the mean extracted energy (Eq. \ref{e:eCmaxwell}).

%{\flushleft\rule{\columnwidth}{1pt}}

%\bibliographystyle{apj}
%\bibliography{MasterRefs8}

\end{document}